\documentclass[a4paper,fleqn,usenatbib]{mnras}

\usepackage{newtxtext,newtxmath}

\usepackage[T1]{fontenc}
\usepackage{ae,aecompl}

\usepackage{graphicx}	
\usepackage{amsmath}	
\usepackage{amssymb}	
\usepackage{color}

\newcommand{\kms}{\,km\,s$^{-1}$} 

\title[High-energy flows in polars]{Numerical simulations of high-energy flows in accreting magnetic white dwarfs}

\author[L. Van Box Som et al.]{
Lucile Van Box Som$^{1,2,3}$\thanks{E-mail: lucile.vanboxsom@cea.fr},
\'E. Falize$^{1,3}$,
J.-M. Bonnet-Bidaud$^{3}$,
\newauthor{M. Mouchet$^{4}$,
C. Busschaert$^{1}$
and A. Ciardi$^{2}$}
\\
$^{1}$CEA-DAM-DIF, F-91297 Arpajon, France\\
$^{2}$LERMA, Observatoire de Paris, PSL Research University, CNRS, Sorbonne Universit\'es, UPMC Univ. Paris 06, F-75005, Paris, France\\
$^{3}$CEA Saclay, DSM/Irfu/Service d'Astrophysique, F-91191 Gif-sur-Yvette, France\\
$^{4}$LUTH, Observatoire de Paris, PSL Research University, CNRS, Universit\'e Paris Diderot, Sorbonne Paris Cit\'e, F-92195 Meudon, France
}

\date{Accepted XXX. Received YYY; in original form ZZZ}

\pubyear{2017}

\begin{document}
\label{firstpage}
\pagerange{\pageref{firstpage}--\pageref{lastpage}}
\maketitle

\begin{abstract}
Some polars show quasi-periodic oscillations (QPO) in their optical light curves which have been interpreted as the result of shock oscillations driven by the cooling instability. Although numerical simulations can recover this physics, they wrongly predict QPOs in the X-ray luminosity and have also failed to reproduce the observed frequencies, at least for the limited range of parameters explored so far. Given the uncertainties on the observed polar parameters, it is still unclear whether simulations can reproduce the observations. The aim of this work is to study QPOs covering all relevant polars showing QPOs. We perform numerical simulations including gravity, cyclotron and bremsstrahlung radiative losses, for a wide range of polar parameters, and compare our results with the astronomical data using synthetic X-ray and optical luminosities. We show that shock oscillations are the result of complex shock dynamics triggered by the interplay of two radiative instabilities. The secondary shock forms at the acoustic horizon in the post-shock region in agreement with our estimates from steady-state solutions. We also demonstrate that the secondary shock is essential to sustain the accretion shock oscillations at the average height predicted by our steady-state accretion model. Finally, in spite of the large explored parameter space, matching the observed QPO parameters requires a combination of parameters inconsistent with the observed ones. This difficulty highlights the limits of one-dimensional simulations, suggesting that multi-dimensional effects are needed to understand the non-linear dynamics of accretion columns in polars and the origins of QPOs.
\end{abstract}




\section{Introduction}

Among the different sources of high-energy processes powered by accretion \citep{Frank2002}, cataclysmic variables (CVs) provide one of the best systems to study the complex conversion of gravitational energy into radiation. Cataclysmic variables are close binary systems composed of a white dwarf accreting matter from a M-type main-sequence star \citep{Warner1995}. Recently, there has been renewed interest in these objects  as potential progenitors of type Ia supernovae \citep{Wheeler2012,Maoz2014}, and whose magnitude-redshift data constrain important cosmological parameters \citep{Riess1998,Perlmutter1999}. To understand the initial conditions of these explosions, it is crucial to determine accurately the properties and the evolution of CVs, and in particular the mass of the white dwarf.

Depending on the magnetic field intensity of the white dwarf, CVs are classified into non magnetic cataclysmic variables for $B_{\text{WD}}< 1$\,MG, intermediate polars (also named DQ Herculis systems) for $B_{\text{WD}}\sim1-10$\,MG, and polars (also named AM Herculis systems) for $B_{\text{WD}}>10$\,MG up to $230$\,MG, the largest value observed so far \citep{Schmidt1999}. In the latter type, the intense magnetic field of the white dwarf locks the whole system into synchronous rotation and prevents the formation of an accretion disk \citep{Cropper1990, Wu2000}. The stream of matter coming from the companion star is captured by the white dwarf magnetic field, it then follows the magnetic lines and falls down onto the magnetic poles of the compact object as an accretion column. This accretion flow strikes the white dwarf photosphere at approximately the free-fall velocity ($\sim 5000$\kms), leading to the formation of a radiative reverse shock. The shock-heated matter is ionized and heated to typical temperatures of $10-50$\,keV and radiates through several processes from the optical domain (optically thick cyclotron radiation) to the X-ray one (optically thin bremsstrahlung radiation). In this high-energy environment, the cooling processes directly depend on white dwarf properties, such as mass, radius and magnetic field as well as the accretion parameters, such as the accretion rate. They shape the density and the temperature profiles of the post-shock region and induce strong gradients near the white dwarf photosphere.

\begin{table*}
\centering
\begin{tabular}{rccccccc}
\hline
\hline
Polars 		& M$_{\text{WD}}$ 		& $B$ 	& $\dot{m}$		&  freq. 	&  opt. 	&  Limit X-ray & Ref.\\
 			& [M$_{\odot}$] 	& [MG] 		& [g\,cm$^{-2}$\,s$^{-1}$] & [Hz] 	&  amp. [\%] 	&  amp. [\%] 			& \\
\hline
AN UMa 		& $1$ 			& $29-36$ 	& $0.2$ 			& $0.62-0.72$ 	& $1.3-3.9$ 			& $25$ 		& (1,2) \\ 
BL Hyi 		& $1$ 			& $12-23$ 	& $1.76$ 		& $0.2-0.8$ 		& $1-4$ 				& $71$ 		& (1,3) \\ 
EF Eri 		& $0.6-0.92$ 			& $16-21$ 	& $6.4$ 			& $0.3-0.6$ 		& $1-1.3$ 			& $58$ 		& (1,4,5) \\
VV Pup 		& $0.73$ 		& $31$ 		& $4.1$ 			& $0.4-1.2$ 		& $1.1-1.7$ 			& $31$ 		& (1,6) \\ 
V834 Cen 	& $0.66-0.85$ 	& $23$		& $0.88-1.4$ 	& $0.26-0.6$		& $0.6-2.1$ 			& $9$ 		& (1,7) \\  
\hline 
\end{tabular}
\caption{Parameters of the five polars showing QPOs with the QPO frequencies (noted freq.) and amplitudes (noted amp.). The limits of X-ray QPO amplitudes are the upper limits derived from the \textit{XMM-Newton} observations. The specific accretion rate is calculated assuming a column cross-section of $10^{15}$\,cm$^{2}$. References : (1) \citet{Bonnet2015}; (2) \citet{Bonnet1996}; (3) \citet{Middleditch1997}; (4) \citet{Cropper1998}; (5) \citet{Larsson1987}; (6) \citet{Ramseyer1993}; (7) \citet{Mouchet2017}.}\label{QPO_data}
\end{table*}

By correctly modelling the emission region, it is possible to determine fundamental properties of the system, such as the mass of the white dwarf. This latter can be derived by fitting synthetic spectra obtained from theoretical models to the observed spectra \citep{Suleimanov2005}. To obtain analytical and semi-analytical solutions from the equations of radiation hydrodynamics which describe the post-shock region, a steady-state regime is frequently assumed \citep{Beuermann2004}. However there are a large number of models of this region including different physical components (radiative processes, gravity, multi-temperature treatment, reflection at the white dwarf surface). Due to the uncertainty of the spatial profiles, the derived value of the white dwarf mass can indeed double or even triple \citep{Cropper1998, Cropper1999}. Consequently the estimation of the white dwarf mass is strongly model-dependent. To assess the validity of the various components of the standard model of accretion columns, the spatial profiles of physical parameters such as temperature or density should be measured. However the small spatial extent ($\sim 100$\,km as derived from the modelling of the post-shock region) prevents direct observations of this area. Moreover, several works \citep{Mauche1998,Girish2007,Ishida2010} have shown that X-ray spectroscopy alone is not sufficient to discriminate between the different models.

Observations of the polars V834 Cen \citep{Middleditch1982}, AN UMa \citep{Middleditch1982}, EF Eri \citep{Larsson1987}, VV Pup \citep{Larsson1989} and BL Hyi \citep{Middleditch1997}, have revealed the presence of excess variability in their optical light curves. These quasi-periodic oscillations (QPOs) are characterized by an amplitude of about $1-5$\% rms and a frequency of about $0.2-1.2$\,Hz which are detailed in Table\,\ref{QPO_data} with the parameters of these polars. These variations of luminosity may be explained by the cooling instability which occurs in the post-shock region due to bremsstrahlung cooling \citep{Langer1981}. Consequently this unsteady effect questions the stability of the post-shock region. The stability properties of radiative shock structures have been investigated considering different approximations such as one temperature flow with a single power-law cooling function of the temperature \citep{Chevalier1982, Imamura1985}, multiple cooling processes \citep{Saxton1998} or two-temperature flows \citep{Saxton2001}. The shock height seems to oscillate up and down with the cooling timescale associated with the bremsstrahlung emission. However these oscillations can be modified very efficiently by cyclotron cooling \citep{Wu2000}. 

Due to these unsteady effects, time-dependent simulations are necessary to compute the non-linear regime of the post-shock region dynamics to obtain relevant information to be linked to observational data. Numerical simulations were performed by \cite{Busschaert2015} to study the dynamics of the accretion column and the evolution of QPOs. Oscillations coupled in the X-ray and optical luminosities were highlighted. Simultaneously a large study of the \textit{XMM-Newton} observations tried to search QPOs in the X-ray luminosity for two dozens of polars \citep{Bonnet2015}. This study showed that no X-ray QPOs were detected at a significant level. Thus these negative results impose upper limits of the X-ray QPO amplitudes.

A recent study \citep{Mouchet2017} has been performed to compare the QPOs in the polar V834 Centuri, observed with the fast ULTRACAM camera mounted on the ESO-VLT, with simulations adapted to the physical parameters of this object. This system is well-known and the only unconstrained parameter is the column cross-section which was varied in the range of $10^{12}-10^{16}$\,cm$^{2}$. The observed QPO amplitude values are compatible with the simulations only in a narrow range of cross-sections around $\sim 4-5\times 10^{14}$\,cm$^{2}$. However the detected frequencies, $0.25-1$\,Hz, are smaller by an order of magnitude than the simulated one, which are in the range of $14-19$\,Hz. There is no cross-section value which can reconcile simulations with observed QPO amplitudes and frequencies. 

This detailed study for one unique polar was unable to provide a consistent interpretation of the oscillations in terms of the shock oscillation model. That is why we propose in this paper to extend the investigation to the five polars showing QPOs. We focus on the evolution of these oscillations as a function of the three key parameters, which are the white dwarf mass, the white dwarf magnetic field and the specific accretion rate. We present new numerical simulations performed with the hydrodynamical RAMSES code \citep{Teyssier2002} where an adapted radiative module has been implemented to model the accretion region physics. In Section \ref{section_1}, the modelling of the accretion column is reviewed through the steady-state regime including gravity. The physics of QPOs is presented, based on the new dynamical simulations in Section \ref{section_2}. Finally, in Section \ref{section_3}, the results of our study are discussed.

\section{Modelling of accretion columns in AM Herculis stars}\label{section_1}

\subsection{The model equations}

\begin{figure}
\centering
\includegraphics[width=5.8cm]{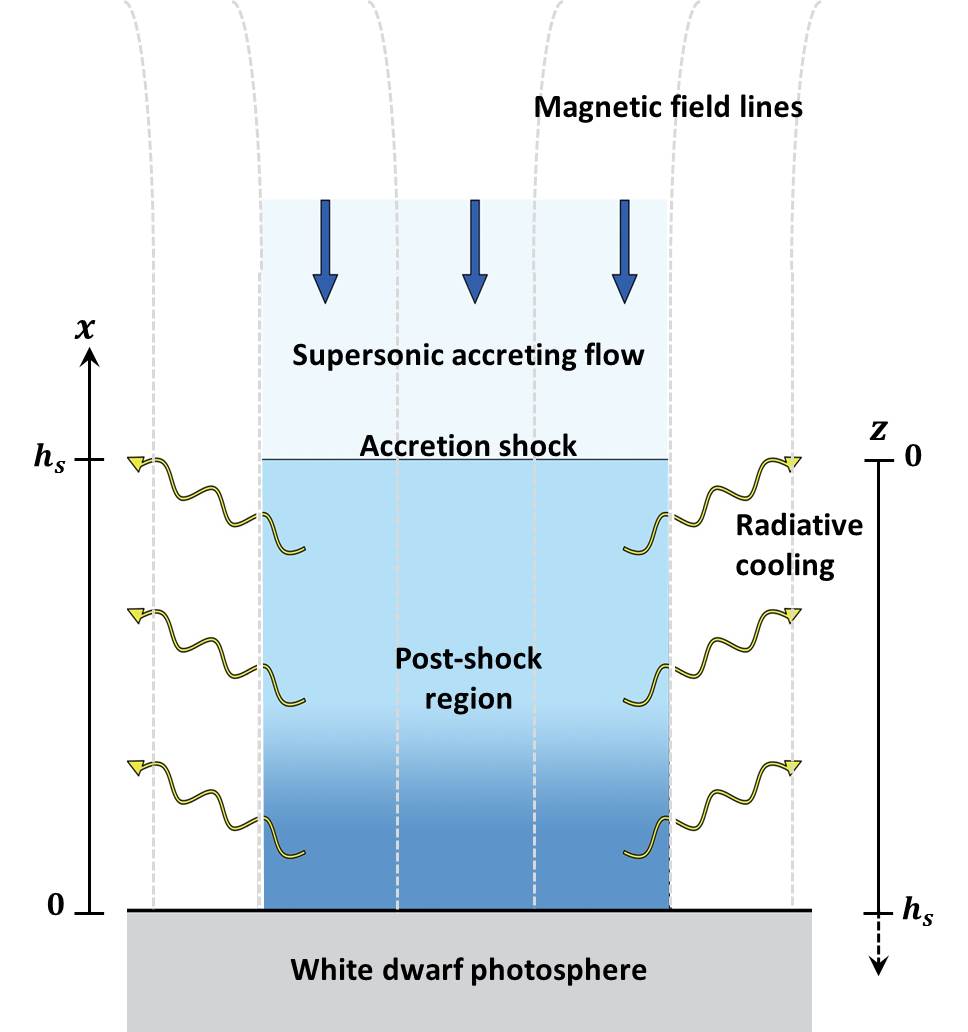}
\caption{Schematic illustration of the accretion column in polars. The supersonic accretion flow hits the white dwarf photosphere. An accretion shock is produced and propagates upstream. In the post-shock region, radiation processes cool matter and shape the density and the temperature profiles. Two different axes are used in this article. The z-axis is used for the resolution of the equations whereas the x-axis is used for figures.}\label{grav_schema}
\end{figure}

The high-energy radiative accretion flow (see Fig. \ref{grav_schema}) is described by the hydrodynamic equations with gravity and radiative cooling. We neglect Compton cooling and conduction in the accretion column \citep{Kylafis1982}. The equations are:
\begin{equation}\label{eq_masse}
\frac{\partial}{\partial t} \rho+ \frac{\partial }{\partial z}(\rho v)=0
\end{equation}
\begin{equation}\label{eq_impulsion}
\frac{\partial }{\partial t}\left(\rho v\right)+ \frac{\partial }{\partial z}\left(P+\rho v^{2}\right)=\rho g_{\text{WD}}
\end{equation}
\begin{multline}\label{eq_energie}
\frac{\partial }{\partial t}\left(\frac{1}{2}\rho v^{2} + \rho \varepsilon \right)+ \frac{\partial }{\partial z}\left(\left[\frac{1}{2} \rho v^{2} + \rho \varepsilon +P\right]v\right)=\\
\rho v g_{\text{WD}} - \Lambda(\rho ,P)
\end{multline}
\begin{equation}\label{EOS}
\varepsilon = \frac{k_{B} T}{\mu m_{H}} \qquad P=(\gamma-1) \rho \varepsilon
\end{equation}
where $t$ and $z$ are respectively the time variable and the spatial coordinate, which takes its origin at the shock front and it is directed towards the white dwarf centre, $\rho$, $v$, $P$, $T$ and $\varepsilon$ are respectively the density, the velocity, the pressure, the temperature and the internal energy per mass unit of the accretion flow, and $g_{WD}$ and $\Lambda$ are respectively the external gravitational field of the white dwarf and the cooling function. Assuming a fully ionized hydrogen gas, the parameter $\gamma$ is the adiabatic index equal to $5/3$. The medium is described as an ideal gas defined by the equation of state equation (\ref{EOS}) where $\mu$ is the mean molecular weight of the gas ($\mu=0.5$ for hydrogen gas), $m_{H}$ is the hydrogen-atom mass, and $k_{B}$ is Boltzmann's constant. 


The gravitational field $g_{\text{WD}}$ of the white dwarf is expressed by:
\begin{equation}\label{gravitation}
g_{\text{WD}} = \frac{G M_{\text{WD}}}{(z_{0}-z)^2}
\end{equation}
where $G$, $M_{\text{WD}}$, and $z_{0}$ are respectively the gravitational constant, the white dwarf mass and a spatial constant defined as $z_{0} = h_{s} + R_{\text{WD}}$, with $h_{s}$ the steady-state height of the shock, and $R_{\text{WD}}$ the white dwarf radius which is linked to the white dwarf mass by the relation given by \citet{Nauenberg}:
\begin{equation}\label{Nauenberg}
R_{\text{WD}} = 7.8\times10^{8}\left[\left(\frac{M_{\text{WD}}}{1.44\,M_{\odot}} \right)^{-2/3}-\left(\frac{M_{\text{WD}}}{1.44\,M_{\odot}} \right)^{2/3}\right]^{1/2}\textrm{\,cm}
\end{equation}
The cooling function $\Lambda$ is given by: 
\begin{equation}
\Lambda = \Lambda_{\text{brem}} + \Lambda_{\text{cycl}}
\end{equation}
where $\Lambda_{\text{brem}}$ and $\Lambda_{\text{cycl}}$ are respectively the bremsstrahlung and the cyclotron processes. Each process can be expressed either exactly or approximatively as a power-law function:
\begin{equation}
\Lambda_{i}=\Lambda_{0,i}\rho^{\alpha_{i}}T^{\beta_{i}}
\end{equation}
where $\Lambda_{0,i}$, $\alpha_{i}$ and $\beta_{i}$ are three characteristic constants specific to the radiative mechanism. With this parametrization, optically thin bremsstrahlung cooling can be exactly expressed as:
\begin{equation}\label{brem}
\Lambda_{\text{brem}} = 5.01\times 10^{6}  \left(\frac{\rho}{10^{-9}\textrm{\,g\,cm}^{-3}}\right)^{2}\left(\frac{T}{10^{8} \textrm{\,K}}\right)^{0.5} \textrm{\,erg\,cm}^{-3}\textrm{\,s}^{-1} 
\end{equation}
while optically thick cyclotron cooling can be approximated by \citep{SaxtonPhD}:
\begin{multline}\label{cycl}
\Lambda_{\text{cycl}}= 7.15\times 10^{7}  \left(\frac{S}{10^{15} \textrm{\,cm}^{2}}\right)^{-0.425} \left(\frac{B}{10 \textrm{\,MG}}\right)^{2.85} \\
\left(\frac{\rho}{10^{-9} \textrm{\,g\,cm}^{-3}}\right)^{0.15}\left(\frac{T}{ 10^8\textrm{\,K}}\right)^{2.50}\textrm{\,erg\,cm}^{-3}\textrm{\,s}^{-1}
\end{multline}   
where $S$ and $B$ are respectively the column cross-section and the magnetic field in the column. The effective cyclotron cooling defined by equation (\ref{cycl}) is deduced from the blackbody emission approximated by the Rayleigh-Jeans relation at low frequencies \citep{Wada1980}.

To estimate the relative importance between the two radiative processes, we define the parameter $\epsilon_{s}$ as the ratio of the characteristic cooling times, just behind the shock front, associated with the bremsstrahlung and the cyclotron cooling. The cooling time is given by $t_{\text{cool}}\sim P/\Lambda$, with $P$ the thermal pressure and $\Lambda$ the cooling function. The parameter $\epsilon_s$ can then be expressed as a function of the polar parameters \citep{Saxton1998} as:
\begin{multline}\label{epsilon_s}
\epsilon_{s} =\frac{t_{\text{brem}}}{t_{\text{cycl}}}= 4.32\,\left(\frac{S}{10^{15} \textrm{\,cm}^{2}}\right)^{-0.425} \left(\frac{B}{10 \textrm{\,MG}}\right)^{2.85} \\
\left(\frac{\dot{m}}{1\textrm{\,g\,cm}^{-2}\textrm{\,s}^{-1}}\right)^{-1.85}
\left( \frac{M_{\text{WD}}}{0.8 \textrm{\,M}_{\odot}} \right)^{2.925} \left( \frac{R_{\text{WD}}}{7\times 10^{8} \textrm{\,cm}} \right)^{-2.925}
\end{multline}
where $\dot{m}$ is the specific accretion rate. Bremsstrahlung emission is dominant when $\epsilon_{s}<1$, whereas cyclotron emission dominates for $\epsilon_{s}>1$. As we shall see later, the dynamics of the shock front depends critically on the relative importance of these two radiative regimes, and thus the value of $\epsilon_{s}$.

To quantify the plasma collisionality and magnetization under typical accretion flow conditions, we compare the ion mean-free-path ($\lambda_{ii}$) and the ion Larmor radius ($r_{Li}$) to the height of the accretion shock ($h_{s}$). 

The ion mean-free-path is given by $\lambda_{ii} = v_{Ti}\tau_{ii}$, where $v_{Ti}=\sqrt{k_{B}T_{i}/m_{i}}$ is the ion's thermal speed and $\tau_{ii}$ is the ion-ion collision time. The latter is given by \citep{Braginskii}:
\begin{equation}
\tau_{ii}= \frac{3\sqrt{m_H}\,T_{i}^{3/2}}{4\sqrt{\pi}e^4 Z^4 \textrm{ln\,}\Lambda\,n_{i} }
\end{equation}
where $T_{i}$, $e$, $Z$, $n_{i}$, ln\,$\Lambda$ are respectively the ion temperature taken equal to the post-shock temperature given by the Rankine-Hugoniot conditions, the electron charge, the
atomic charge state (here $Z=1$), the ion particle density and the Coulomb logarithm. For typical densities and temperatures in the post-shock region of the polars displaying QPOs (Table\,\ref{QPO_data}), the Coulomb logarithm is $\sim 17$ and the ordering of the characteristic lengths is:
\begin{equation}
h_{s}\sim10^6-10^7\textrm{cm} \gtrsim \lambda_{ii}\sim 10^{4}-10^{6}\textrm{cm} \gg r_{Li}\sim10^{-4}-10^{-3}\textrm{cm}
\end{equation}
For most of the polar parameters ($M_{\text{WD}}\lesssim 1$ M$_{\sun}$ and $B\lesssim 30$ MG), the ion scattering lengths due to Coulomb collisions are short and the accretion flow is then collisional. However we note that for larger masses and magnetic fields ($M_{\text{WD}} \gtrsim 1$ M$_{\sun}$ and $B\gtrsim 30$ MG), the mean-free-path may be longer than the characteristic length scale of the accretion columns: $l_{ii}\sim h_{s}$. In those cases, the very short ion Larmor radius however justifies the localization perpendicular to the magnetic field, while along the field lines plasma microfluctuations and magnetic entanglement provide the necessary scattering. 

To justify the equilibration between ion and electron temperatures ($T_{e}\simeq T_{i}$), we compare the electron-ion equilibration time ($\tau_{ei}$) with the cooling time ($t_{\text{cool}}$). The characteristic time of energy exchanged between electrons and ions is defined by \citep{Spitzer}:
\begin{equation}
\tau_{ei} =  \frac{3 m_{e}m_{H}k_{B}^{3/2}}{8(2 \pi)^{1/2}n_{i} Z^2 e^4 \textrm{ ln\,}\Lambda} \left(\frac{T_{e}}{m_{e}}+\frac{T_{i}}{m_{H}}\right)^{3/2}
\end{equation}
To estimate a maximum energy exchange time, $T_{i}$ is taken equal to the post-shock temperature given by the Rankine-Hugoniot conditions and $T_{e}$ is taken equal to the pre-shock temperature assumed at $10^6$\,K. Thus for most of the polar parameters investigated, $\tau_{ei}/t_{\text{cool}}\sim 10^{-6}-10^{-1}$ which confirms that the temperature equilibrium can describe the post-shock flow in accretion columns. 

Consequently, the accretion flows can be studied using fluid-like equations with a single temperature model defined by the equations\,(\ref{eq_masse}), (\ref{eq_impulsion}), and (\ref{eq_energie}).

\subsection{The steady-state solutions}

Analytical and semi-analytical solutions of equations (\ref{eq_masse}), (\ref{eq_impulsion}), and (\ref{eq_energie}) can be found assuming steady-state.
These solutions are used to estimate the mass of the white dwarf by fitting the synthetic X-ray spectrum to the observed one \citep{Suleimanov2005} and they are useful to validate numerical simulations. As we shall see, the complete semi-analytical solutions can be classified into two regimes according to gravity. Its importance can be quantified by the ratio between the steady-state height and the white dwarf radius, $h_{s}/R_{\text{WD}}$. Our simulations show that gravity can be neglected for $h_{s}/R_{\text{WD}}\lesssim 0.01$, whereas it is essential to include gravitational effects when $h_{s}/R_{\text{WD}}\gtrsim 0.01$ to correctly describe the accretion flow.

To solve the steady-state system of equations (\ref{eq_masse}), (\ref{eq_impulsion}), and (\ref{eq_energie}), we introduce the inverse of the density contrast, noted $\eta$  \citep{Falize2009_rad}:
\begin{equation} \label{eta}
\rho(z)=\frac{\rho_{0}}{\eta(z)} , \qquad
v(z)=v_{0}\eta(z)
\end{equation}
where $\rho_{0}$ and $v_{0}$ are respectively the density and the velocity of the supersonic, upstream flow, which is given by the free-fall velocity at the position of the steady-state shock front:
\begin{equation}\label{vs_grav}
v_{0} = \sqrt{\frac{2G M_{\text{WD}}}{z_{0}}}
\end{equation}
and the upstream density is inferred from the specific accretion rate, $\dot{m}$, as $\rho_{0} = \dot{m}/v_{0}$.

After some algebraic manipulations, equations (\ref{eq_masse}), (\ref{eq_impulsion}) and (\ref{eq_energie}) are reduced to two ordinary differential equations coupling the variables $\eta$ and $P$:
\begin{align}\label{eq_grav}
\frac{dP}{dz} &=  \frac{\rho_{0}}{\eta}\frac{GM_{\text{WD}}}{(z_{0}-z)^{2}}-\rho_{0}v_{0}^{2}\frac{d\eta}{dz}\\
\frac{d\eta}{dz}\left[\gamma P v_{0} - \rho_{0}v_{0}^{3} \eta \right] &= - (\gamma-1)\Lambda(\eta, P) -v_{0}\rho_{0}\frac{GM_{\text{WD}}}{(z_{0}-z)^{2}}\label{eq_grav2}
\end{align}
We note that the transformation defined by equation (\ref{eta}) allows the reduction of the system (\ref{eq_masse}), (\ref{eq_impulsion}), and (\ref{eq_energie}) in a more direct way than in previous studies \citep{Cropper1999}. The system (equations (\ref{eq_grav}) and (\ref{eq_grav2})) is solved by a dichotomy method coupled with a Fourth-order Runge-Kutta method assuming the Rankine-Hugoniot conditions at the shock front and a cold and solid wall, where temperature and velocity reach zero, at the white dwarf photosphere. The shock height is extracted iteratively from the boundary condition: $v = 0$ at the white dwarf photosphere. 

\begin{figure}
\centering
\includegraphics[width=9cm]{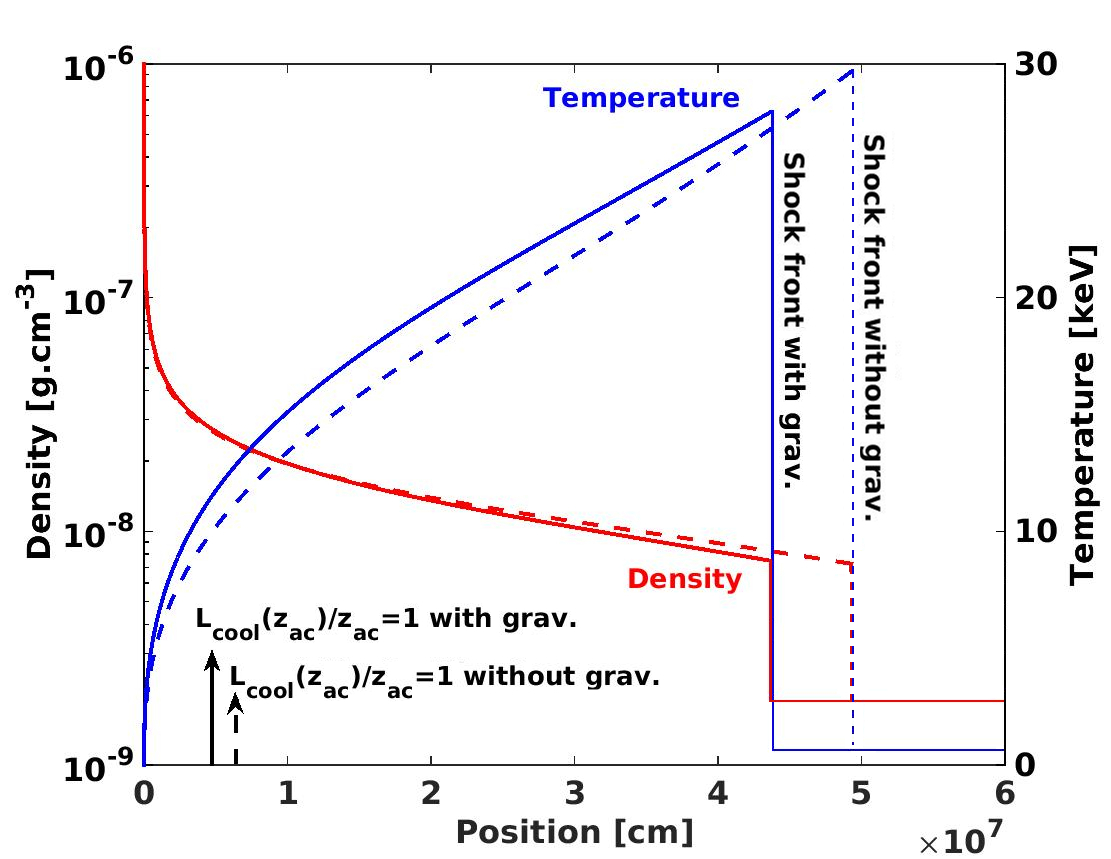}
\caption{Density (red lines) and temperature (blue lines) profiles of the post-shock region for $M_{\text{WD}} = 0.8$\,$M_{\odot}$, $B=10$\,MG and $\dot{m}=1$\,g\,cm$^{-2}$\,s$^{-1}$. The profiles in dotted lines are the solutions without gravity whereas the profiles in full lines are the solutions with gravity. The white dwarf photosphere is located at the origin of the axis and the shock front is located at the vertical lines. The position of the ratio $L_{\text{cool}}(z_{\text{ac}})/z_{\text{ac}}=1$ is indicated for both cases.}\label{grav_profils}
\end{figure}

The steady-state temperature and density profiles are displayed in Fig.\,\ref{grav_profils} by respectively the blue and red lines for $M_{\text{WD}} = 0.8$\,$M_{\odot}$, $B=10$\,MG and $\dot{m}=1$\,g\,cm$^{-2}$\,s$^{-1}$. By taking gravity into account in the equations, the steady-state shock height is reduced from $4.93\,\times\,10^7$\,cm to $4.36\,\times\,10^7$\,cm, which implies about $10\%$ of reduction. Then $h_{s}/R_{\text{WD}}\sim 0.06$. However the modifications in the profiles are more important for temperature than density. The shock temperature is lower with a flatter profile but the post-shock region is significantly hotter due to the release of gravitational energy in agreement with the results presented by \citet{Cropper1999}. The gravity model induces harder spectra than the one without it. Thus fitting observed spectra with models without gravity will tend to overestimate the white dwarf mass. We point out that cyclotron cooling also tends to decrease the height of the accretion column, as described by previous works \citep{Wu1994, Wu2000}, because it reduces the post-shock temperature profile and makes the post-shock region denser.

When gravity is negligible ($h_{s}/R_{\text{WD}}<0.01$), the system composed of equations\,(\ref{eq_grav}) and (\ref{eq_grav2}) can be simplified to a single differential equation which can be solved by a Fourth-order Runge-Kutta method. Some analytical solutions can be found when only one emission process dominates in the post-shock region, and when it may be expressed as a power-law function of the type $\Lambda \sim  \rho^\delta P^\zeta$ \citep{Chevalier1982, Laming2004}. These solutions are particularly useful to obtain an estimate of the steady-state height as a function of the polar parameters.

In this particular case, an implicit function of $\eta$ is given by:
\begin{multline}\label{tot_2F1}
\gamma\frac{\eta(z)^{\delta+1}}{\delta+1} \left[1+\frac{1}{\gamma M^{2}}\right]^{1-\zeta} \times _ {2}F_{1}\left(\zeta,\delta+1;\delta+2;\frac{\eta(z)}{1+1/[\gamma M^{2}]}\right)- \\
(\gamma+1)\frac{\eta(z)^{\delta+2}}{\delta+2}\left[1+\frac{1}{\gamma M^{2}}\right]^{-\zeta} \times 
 _2F_1\left(\zeta,\delta+2;\delta+3;\frac{\eta(z)}{1+1/[\gamma M^{2}]}\right)  \\
 =-( \gamma-1)\Lambda_{0}\rho_{0}^{\delta+\zeta-1} v_{0}^{2 \zeta-3}z + C
\end{multline}
where $_{2}F_{1}$ are Gauss hypergeometric functions \citep{Abra}, $M$ is the Mach number and the constant \textit{C} is defined by the value of equation (\ref{tot_2F1}) at the shock front in the Rankine-Hugoniot conditions. 
For specific values of $\delta$ and $\zeta$, such as for the bremsstrahlung cooling ($\delta=1.5$ and $\zeta=0.5$), equation (\ref{tot_2F1}) can be simplified in usual functions and recovers the published analytical solutions. For a hydrogen plasma ($\gamma=5/3$) and in the strong shock approximation with bremsstrahlung cooling, an analytical steady-state shock height can be calculated. This solution can be expressed as a function of the specific accretion rate and the white dwarf mass and radius \citep{Wu1994} :
\begin{equation}\label{xs_stat}
h_{s} = 10^{8} \textrm{\,cm}\left(\frac{\dot{m}}{1\textrm{\,g\,cm}^{-2}\textrm{\,s}^{-1}}\right)^{-1}\left(\frac{M_{\text{WD}}}{0.8 \textrm{\,M}_{\odot}}\right)^{3/2}\left(\frac{R_{\text{WD}}}{7\times10^{8} \textrm{\,cm}}\right)^{-3/2}
\end{equation}
Equation \,(\ref{xs_stat}) gives an upper-limit of the steady-state shock height as a function of the polar parameters because under more general conditions the shock height decreases when gravity and cyclotron cooling are present.



\section{Time-dependent dynamics of the accretion flow}\label{section_2}

Five polars have shown quasi-periodic oscillations (QPOs) in their optical light curves. The polars parameters and the QPO characteristics are summarized in Table (\ref{QPO_data}). Due to these unsteady phenomena, time dependent simulations of the accretion flow are carried out to explore the non-linear dynamics of the post-shock region.

\subsection{Radiative instabilities}

According to the first law of thermodynamics, the internal energy of a radiative flow is expressed as $de=\delta W + \delta Q$ where $\delta W$ is the amount of work equal to $\delta W = -PdV$ and $\delta Q$ is the cooling losses expressed as $\delta Q = -\left(\Lambda /\rho\right)  dt$ \citep{Boily1995}. If heating due to compression is more effective than cooling ($de/(-PdV) >0$), the flow is then stable. When cooling is strongly dominant ($de/(-PdV)  < 0$), it is consequently unstable. When the cooling function is characterized by a power-law expression: $\Lambda \sim \rho^2 T^\beta$, the cooling instability can develop in a radiative flow if $\beta <1$ \citep{Field1965}. 

 
In the classical model of accretion columns \citep{Langer1981}, the QPOs result from a combination of bremsstrahlung cooling trying to collapse the post-shock region and the resistance of the post-shock flow to being compressed. In this model, there is no physical process which prevents the collapse of the column. Thus the oscillations are resulting from the non-linear nature of the cooling instability. However, previous work \citep{Busschaert2015} has already numerically shown that these QPOs in accretion columns are induced by the collisions between the accretion shock and secondary shocks triggered by a second radiative instability. The latter was already seen in previous simulations but its role was not recognised \citep{ Strickland1995,Mignone2005}. In this paper, we confirm this mechanism which implies a subtle interplay between shocks to sustaining the oscillations at the average height predicted by steady-state accretion models. Besides we demonstrate that these secondary shocks are coming from an acoustic horizon near the white dwarf photosphere, triggered by a second radiative instability.

In this case, a radiative shocked flow can become supersonic again if the cooling is such that sound velocity decreases faster than the flow velocity. When the cooling function is a power-law function, it has to be characterized by $\beta<3/2$ \citep{Falle1981} to develop the secondary shock instability. This condition is indeed fulfilled by bremsstrahlung cooling within the accretion columns of polars. It means that the characteristic local cooling length should be smaller than the steady-state shock height. 

As in \citet{Bertschinger1989}, we define in the post-shock region the local characteristic cooling length, as the product of the local sound speed, $c_{s}(z)$, and the local cooling time, $t_{\text{cool}}(z)$:
\begin{equation}\label{Lcool}
L_{\text{cool}}(z) = c_{s}(z)\times t_{\text{cool}}(z)
\end{equation}
where $z$ is defined by Fig.\,\ref{grav_schema} with its origin at the shock front. The value of the ratio $L_{\text{cool}}(z)/z$ is maximum near the shock front and decreases towards the photosphere of the white dwarf. In the cooling regimes relevant to polars, this length is always smaller than the height of the accretion column, $h_s$. Thus an acoustic horizon arises at position $z_{\text{ac}}$ where the local cooling length is equal to the distance to the shock front: $L_{\text{cool}}(z_{\text{ac}})/z_{\text{ac}}=1$. According to the semi-analytical solutions, the acoustic horizon position is about $10\%$ of the steady-state shock height above the white dwarf photosphere as shown in Fig.\,\ref{grav_profils}. As a result, and as shown later, we expect the secondary shock to appear at this point. However semi-analytical solutions are not sufficient and time-dependent simulations are necessary to study the interplay between the two shocks triggered by radiative instabilities. 

\subsection{Numerical set-up}

\begin{table*}
\centering
\begin{tabular}{rcccccc}
\hline
\hline
Models 		& M$_{\text{WD}}$ 	& $B$ 	& $\dot{m}$ & freq. 	 	& $L_{X}$ amp. & $L_{\text{opt}}$ amp. \\
			& [M$_{\odot}$] & [MG] 	& [g\,cm$^{-2}$\,s$^{-1}$] & [Hz] & [\%] & [\%] \\
\hline
(a) & $0.5$ 	& $0-22$ 	& $1$ 		& $1.6-15$ 	& $0.1-50$ 	& $0.1-30$   	\\
(b) & $0.8$ 	& $0-11$ 	& $1$ 		& $0.8-9$ 	& $0.1-53$ 	& $0.1-20$  	\\ 
(c) & $1$ 	& $0-4$ 		& $1$		& $0.5-2$ 	& $0.1-60$ 	& $0.1-20$  	\\ 
(d) & $0.5$ 	& $20$ 		& $0.8-100$ 	& $9-151$ 	& $0.1-50$ 	& $0.1-30$   	\\ 
(e) & $0.8$ 	& $20$ 		& $3-100$ 	& $6-68$ 	& $0.1-53$ 	& $0.1-26$   	\\  
(f) & $1$ 	& $20$ 		& $7-200$ 	& $10-87$ 	& $0.1-54$ 	& $0.1-30$   	\\ 
\hline 
\end{tabular}
\caption{Parameters and results of the numerical models. The chosen parameters enclose the range of values displayed by the five QPO polars presented in Table \ref{QPO_data}. The amplitudes are relative to the mean value of the luminosities. The parameter space is covered by about one hundred simulations.}\label{liste_simu}
\end{table*}

We solve equations (\ref{eq_masse}), (\ref{eq_impulsion}), and (\ref{eq_energie}) with the RAMSES code \citep{Teyssier2002} in which a radiative module has been implemented and the white dwarf gravitational potential has been integrated. The implemented function of the cooling processes is 
written in the form given by \citet{Wu1994}: 
\begin{multline}\label{Lambda_tot}
\Lambda (\rho, P ) = \Lambda_{0}^{\text{brem}} \rho ^{1.5} P^{0.5} \\
\left[ 1+ \epsilon_{s}\left(M_{\text{WD}},B,S,\dot{m}\right)\left( \frac{P}{0.75 \rho_{0} v_{0}^2}\right)^2 \left(\frac{4\rho_{0}}{\rho}\right)^{3.85}\right]
\end{multline}
where $\Lambda_{0}^{\text{brem}}=3.9\times 10^{16}$\,g$^{-1}$\,cm$^{4}$\,s$^{-2}$, $\rho_{0}$ and $v_{0}$ are respectively the density and the velocity of the supersonic-upstream flow, and $S$ is the column cross-section. In all simulations, $S$ is assumed constant at $10^{15}$\,cm$^{2}$. This hypothesis will be discussed later. Since the Mach number of the incoming flow is around $30$, a HLLC (Harten-Lax-van Leer Centrale wave) Riemann solver is used to capture shocks and solve the high Mach number flow \citep{Toro}.  

At the upper boundary of the computational domain, Dirichlet conditions are used to model the homogeneous incident flow from the low-mass companion star characterized by the density, $\rho_{0}$, of the incoming flow and the velocity $v_{s}$. At the other side, a specific boundary condition has been implemented to mimic accretion onto the white dwarf photosphere modelled by the following equation:
\begin{equation}\label{condition_bord}
[\rho v]^{\text{GR}} = [\rho v]^{\text{Fl}} \left[1-2\left(\frac{\rho_{0}}{[\rho]^{\text{Fl}}}\right)\right]
\end{equation}
where $[\rho v]^{\text{GR}}$ is the mass flux in the ghost region, $[\rho v]^{\text{Fl}}$ and $[\rho]^{\text{Fl}}$ are respectively the mass flux and the density in the computational box. This condition conserves the mass and the energy between the computational domain and the ghost region. At the first iteration of time, the white dwarf photosphere behaves as a perfect wall because $[\rho]^{\text{Fl}}=\rho_{0}$ in equation (\ref{condition_bord}) and $[\rho v]^{\text{GR}}=-[\rho v]^{\text{Fl}}$. Since radiative losses increase the accreted matter density near the white dwarf photosphere, the photosphere condition absorbs gradually the accreted mass until it achieves the asymptotically steady-state regime defined by $[\rho v]^{\text{GR}}=[\rho v]^{\text{Fl}}$. A comparison between our simulations and the simulations performed by \citet{Busschaert2015} have shown that the dynamics of the post-shock is in good agreement despite the difference of boundary conditions.

The simulations are performed using a 1D uniform Cartesian grid without the available Adaptive Mesh Refinement (AMR) present in RAMSES. For all the simulations presented here, the resolution of the grid cells has been optimized to provide an accuracy in the range $0.1-1$\,km. Typically the post-shock region is resolved with a thousand cells. A summary of simulations are displayed in Table\,\ref{liste_simu} for different polar parameters. We have selected about the mean value of the white dwarf mass of the polars showing QPOs ($M_{\text{WD}} = 0.8$\,M$_{\sun}$) and the two extrema to include all possible white dwarf masses ($M_{\text{WD}} = 0.5$ and $1$\,M$_{\sun}$). For the three different white dwarf masses, the accretion rate and the magnetic field are varying respectively, as much as over the range $0.8-200$\,g\,cm$^{-2}$\,s$^{-1}$ and $0-22$\,MG. The parameter space is covered by about one hundred simulations. We have also run an extensive set of simulations for larger magnetic fields and smaller accretion rates, where cyclotron cooling dominates. As expected from linear theory of the cooling instability \citep{Wu2000}, the simulations show that the oscillations are completely inhibited and that the column is stable.

\subsection{Shock front dynamics}

\begin{figure}
\centering
\includegraphics[width=9cm]{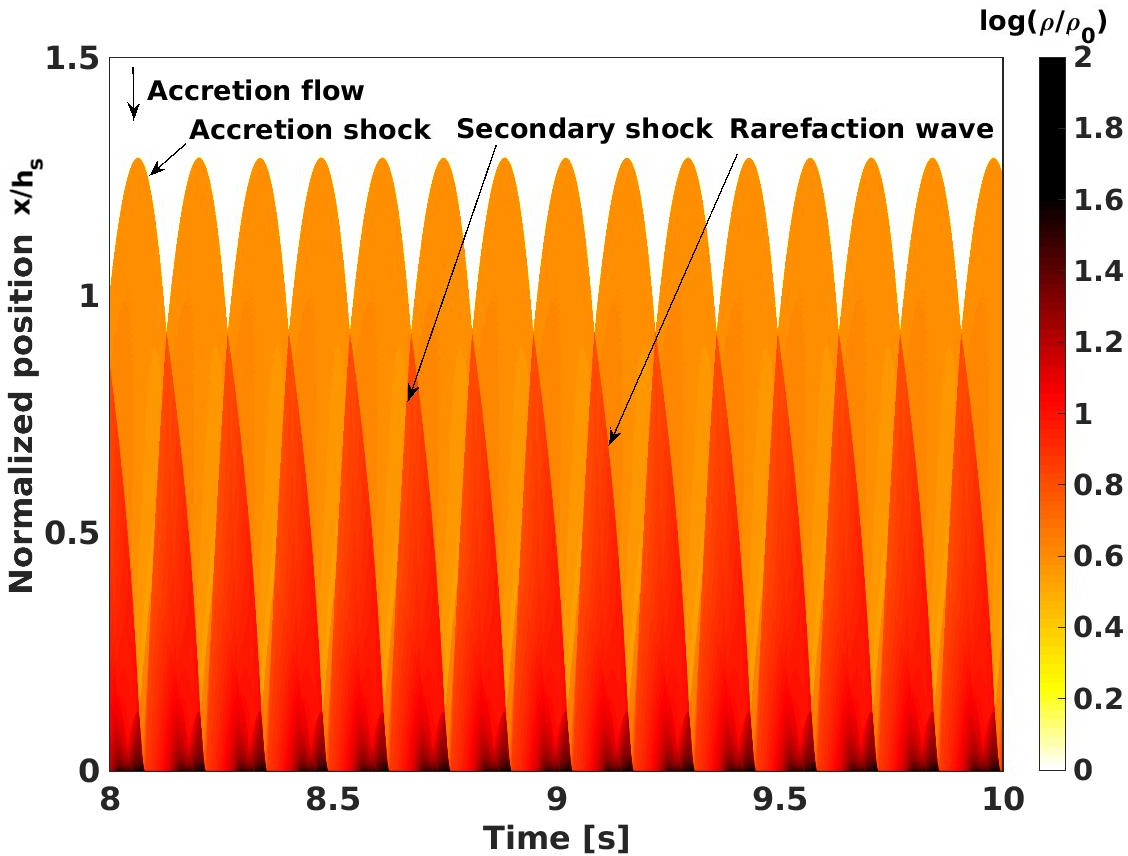}
\caption{Space-time diagram of the accretion flow for a $0.8\,$M$_{\odot}$ and $10\,$MG white dwarf and with a specific accretion rate of $10$\,g\,cm$^{-2}$\,s$^{-1}$. The incoming flow is coming from the top and the white dwarf photosphere is located at the bottom ($x=0$). The coloured scale shows the logarithm density normalized by the initial density. In this figure, the complex dynamics of the radiative region is illustrated.}\label{space_time}
\end{figure}

To better understand the shock front dynamics, the structure and the evolution of the post-shock region are presented on a space-time diagram in Fig.\,\ref{space_time}. We have chosen the polar parameters to illustrate the development of the cooling instability and the secondary shock. They are characterized by a white dwarf mass of $M_{\text{WD}}=0.8$\,M$_{\odot}$, a magnetic field of $10$\,MG and  a specific accretion rate of $\dot{m}=10$\,g\,cm$^{-2}$\,s$^{-1}$. The magnetic field and the accretion rate are chosen to have a dominant bremsstrahlung cooling in the post-shock region with a ratio between the two radiative processes equal to $\epsilon_{s} = 0.058$. The coloured scale shows the logarithm density normalized by the initial density. The accretion flow is coming from the top with a density of $\rho_{0}$ in white. The white dwarf photosphere is located at the bottom ($x=0$). The shock front oscillates around its steady-state value ($h_{s}=9.85\times 10^{6}$\,cm). Transient phenomena associated with the intial formation of the accretion shock are observed at early times ($\lesssim 8$\,s) and are not displayed in Fig.\,\ref{space_time}.

To sustain the accretion shock oscillations at the steady-state height, an internal structure is setting up with a complex wave and shock structure. The location of the acoustic horizon defined in the previous section is calculated at each time step and corresponds to the creation zone of the secondary shock. The latter is created at about $10\%$ of the shock height above the white dwarf photosphere as expected by the steady-state profiles. It then propagates upstream up to the main accretion shock until the two shocks collide. After the collision, the main shock propagates counter the accretion flow and a rarefaction wave is created which travels down to the white dwarf photosphere as shown in Fig.\,\ref{space_time}. After a few moments, a cycle is established between the two shocks with a merging of these two structures. This limiting cycle continues until the end of the simulations without any perturbation. The secondary shock oscillates, at the same frequency as the shock front in the shock front referential. The secondary shock is clearly identified in the post-shock region in Fig.\,\ref{space_time}. The secondary shock Mach number is around $1.2-1.7$ which is in a good agreement with the published values by \citet{Busschaert2015}.

\begin{figure}
\centering
\includegraphics[width=8.9cm]{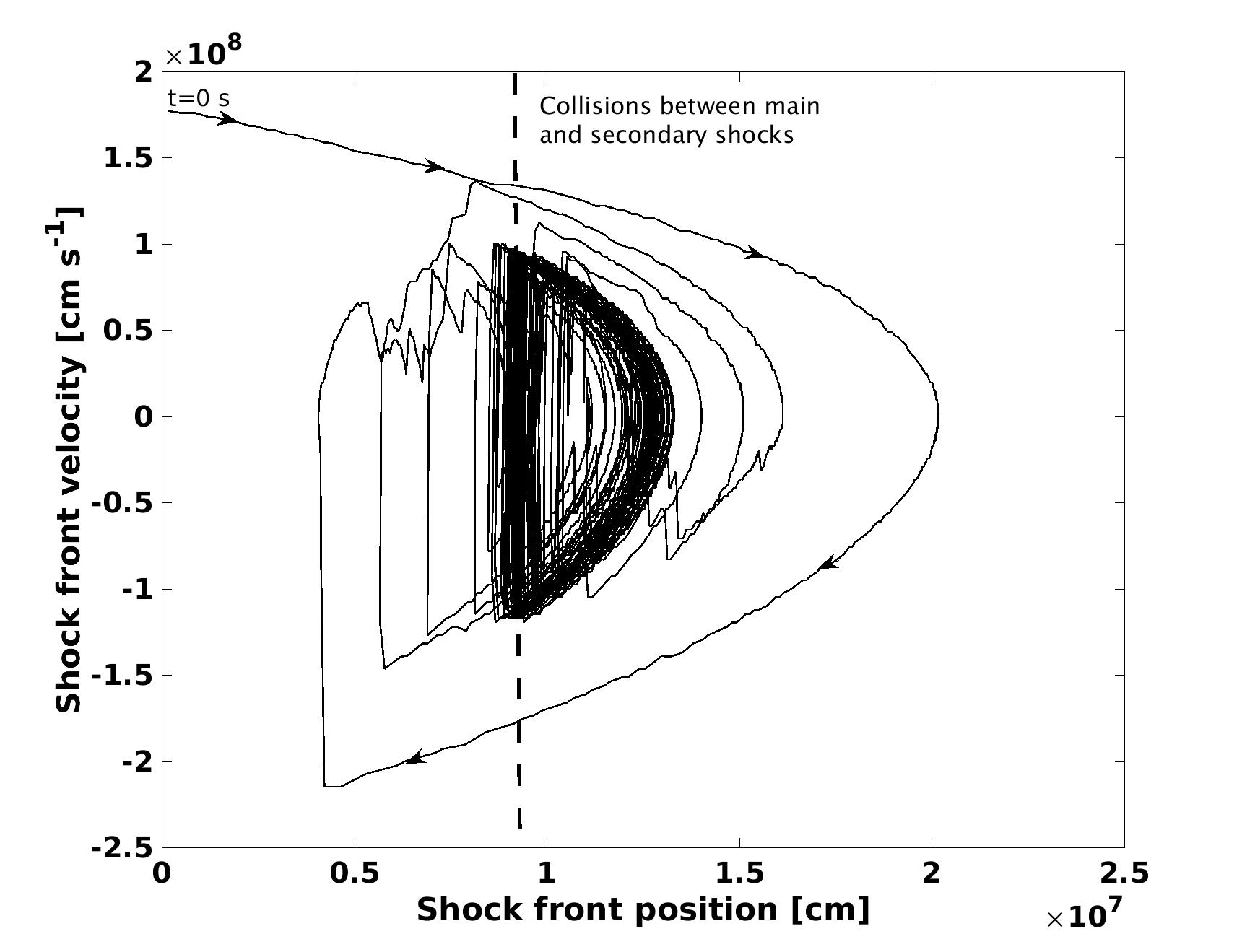}
\caption{Phase diagram of the shock front: velocity as a function of the normalized position of the shock front for the same parameters as in Fig. \ref{space_time}. The \textit{D} structure is achieved after about $3$\,s. The collisions between main and secondary shocks induce an inversion of the shock front velocity displayed by the vertical dotted line.}\label{diagram_phase}
\end{figure}

To identify the collisions between the two shocks, the evolution of the shock front velocity as a function of its normalized position is presented in Fig.\,\ref{diagram_phase}. At the beginning, the dynamics in the phase-diagram is a clockwise convergent spiral which very rapidly winds up to form a \textit{D} structure after about $3$\,s. The vertical bar is due to the collision between the two shocks which change suddenly the velocity of the shock front. The clockwise direction is obviously constant whatever the polar parameters. The \textit{D} pattern observed in the phase-diagram is characteristic of the interaction between the main accretion shock and the secondary shock, and it clearly illustrates the saturation of the amplitude of the oscillation of the accretion shock. The secondary shock is thus an essential part of the shock front dynamics and structure: its presence halts the collapse of the accretion shock onto the white dwarf photosphere, maintaining it at an average height which corresponds to the steady-state height $h_s$. Simulations of accretion columns in the Young Stellar Objects, where no secondary shocks are observed, display indeed the collapse of the accretion shock onto the stellar surface \citep{Orlando2013}. Therefore, in polars, oscillations are driven by a subtle interplay between shocks triggered by cooling instabilities.

\subsection{Luminosities: probes of QPOs}

To better link the numerical simulations to astronomical observations, synthetic X-ray, $L_{X}$,  and optical, $L_{\text{opt}}$, luminosities are calculated. As X-ray radiation is mainly due to bremsstrahlung emission whereas optical radiation is due to  cyclotron emission, the luminosities are respectively defined by:
\begin{equation}\label{luminosX:eq}
L_{X}(t) = S \int^{z_{0}}_{R_{\text{WD}}}\Lambda_{\text{brem}}' dz
\end{equation} 
\begin{equation}\label{luminosO:eq}
L_{\text{opt}}(t) = S \int^{z_{0}}_{R_{\text{WD}}} \Lambda_{\text{cycl}} dz
\end{equation} 
where $S$ is the column cross-section assumed constant at $10^{15}$ cm$^{2}$, $\Lambda_{\text{brem}}'$ is the bremsstrahlung cooling integrated in the X-ray domain and  $\Lambda_{\text{cycl}} $ is the cyclotron cooling defined by equation (\ref{cycl}). 
To connect with the \textit{XMM-Newton} observations, $\Lambda_{\text{brem}}'$ is integrated over the range of detections $[0.5- 10\,$keV$]$ as in \citet{Bonnet2015}:
\begin{multline}
\Lambda_{\text{brem}}' = \int^{10}_{0.5} 2.44\times 10^{10}\textrm{  } [\rho(x,t)]^{2}\textrm{  } [T(x,t)]^{-0.5} \\
g_{B}(E,T(x,t))\times\textrm{ exp}\left(-{E}/{k_{B} T(x,t)}\right) dE
\end{multline} 
where $E$ and $g_{B}$ are respectively the emitted photon energies and the Gaunt factor. In high-energy regimes relevant to polars, the Gaunt factor is approximated by $g_{B}(E,T(x,t))\sim \left[{E}/{k_{B} T(x,t)}\right]^{-0.4}$ where $E \sim k_{B}T$ \citep{Zombeck}. 

To determine the main oscillation frequencies, we realized standard Fast Fourier Transform (FFT) of the luminosities and the position of the accretion shock. The simulation data have a resolution time of $t_{\text{cool}}/10$ s and the FFTs have $4096$ data points. The FFTs are calculated from few seconds after the beginning of the simulations to avoid disturbances linked to the initial creation of the accretion shock.

\begin{figure}
\centering
\includegraphics[width=8.8cm]{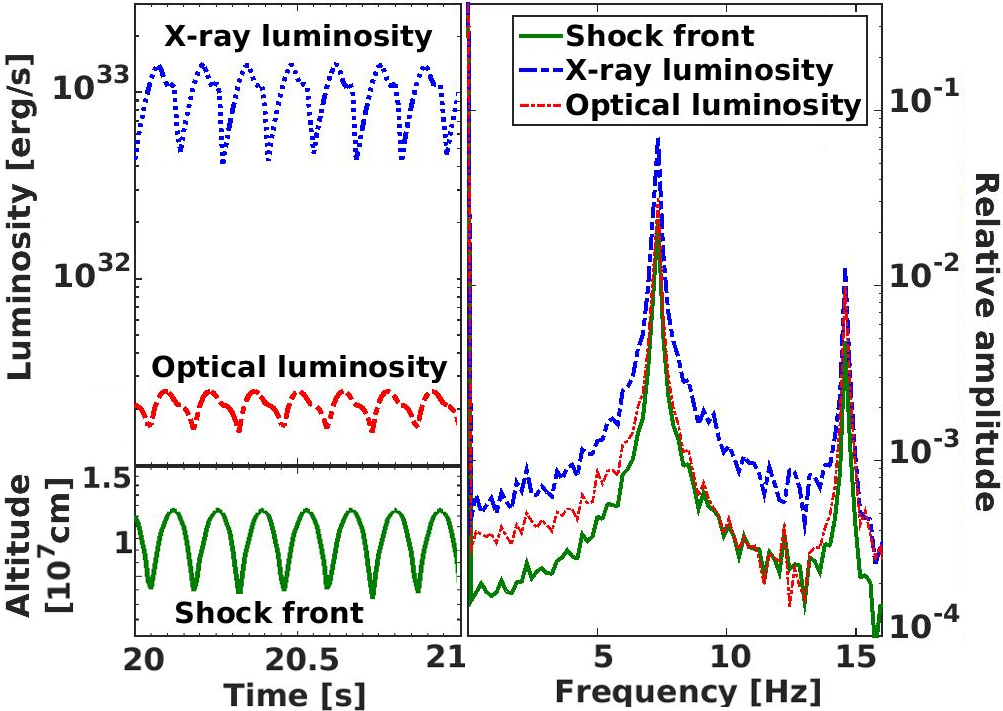}
\caption{Time evolution (at left) and FFTs (at right) of the oscillations of the shock front (green line), the X-ray (blue line) and the optical (red line) luminosities for a $0.8\,$M$_{\odot}$ white dwarf mass, a $10\,$MG magnetic field and a specific accretion rate of $10$\,g\,cm$^{-2}$\,s$^{-1}$ as in Fig. \ref{space_time}. In the FFTs, luminosities are normalized by their mean values.}\label{oscillations}
\end{figure}

\begin{figure}
\centering
\includegraphics[width=8.8cm]{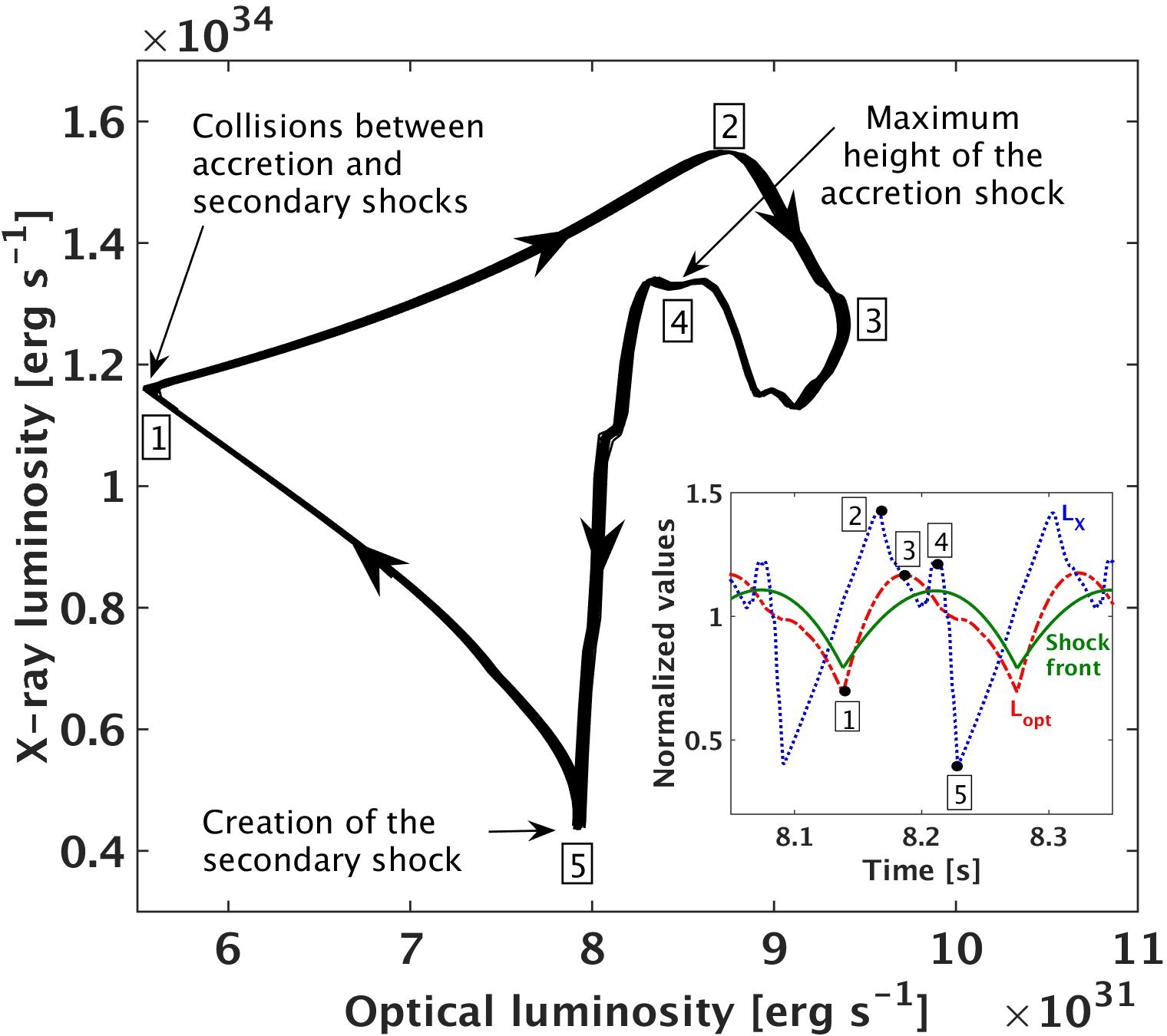}
\caption{Phase diagram of the X-ray luminosity versus the optical luminosity for the same parameters as in Fig. \ref{space_time} and in Fig. \ref{oscillations}. This is a superposition of time between $8$\,s until $25$\,s. This specific pattern illustrates the limit cycle of the luminosities in the post-shock region.}\label{diagram_phase_Lum}
\end{figure}

In Fig.\,\ref{oscillations}, the X-ray (blue line) and optical luminosities (red line) oscillations are displayed in parallel with the shock front oscillations (green line). The shock front oscillations and the optical luminosity are in phase, whereas oscillations in the X-ray luminosity are not synchronized, with a phase difference between peaks of about $\pi/6$. A complex structure between luminosities can be seen in Fig.\,\ref{diagram_phase_Lum} which displays the phase diagram of the X-ray and the optical luminosities. This figure is a superposition of values from $t=8$\,s to  $t=25$\,s. It illustrates the limiting cycle achieved by the post-shock region, which is similar to that seen in Fig\,\ref{diagram_phase}. The particular luminosities shape is due to the strong gradients created in the post-shock region. Indeed the secondary shock dynamics induces density and temperature gradients which modify the luminosities. The collisions between the accretion and the secondary shocks occur at the minimum of the optical luminosity (label number $1$) whereas the creation of the secondary shocks occurs at the minimum of the X-ray luminosity (label number $5$). This very specific pattern appears when the bremsstrahlung cooling is dominant in the post-shock region (i.e. $\epsilon_{s} \ll 1$) as in Fig.\,\ref{diagram_phase_Lum}. On the contrary, when cyclotron cooling is dominant (i.e. $\epsilon_{s} \sim 1$), this pattern disappears and becomes a non-structured full ellipse. In the limiting case of $\epsilon_{s} \gg 1$, when oscillations are completely inhibited, it becomes a point.

As shown by the FFT in Fig.\,\ref{oscillations}, the luminosities are strongly coupled and oscillate at the same frequency as the shock front, $f= 7.2\,$Hz. This frequency is of the order of the inverse of cooling time, $1/t_{\text{cool}}$, which confirms that these oscillations are created by the radiative instabilities in agreement with linear stability analyses \citep{Chevalier1982}. In this context, the secondary shock oscillates in the shock front referential at the same frequency, $f= 7.2$\,Hz, which supports the creation of a limit cycle between the primary and the secondary shocks.

The ratio between X-ray and optical luminosities is an indication of which radiative cooling process is dominant. Indeed in the bremsstrahlung-dominant regime ($\epsilon_{s} \ll 1$), we have that $\langle L_{X}\rangle \gg \langle L_{\text{opt}} \rangle$ as illustrated by Fig.\,\ref{oscillations}. The mean value of the luminosities in Fig.\ref{oscillations}, noted $\left\langle L\right\rangle$, is calculated by integrating the luminosities over the total simulation time : $\left\langle L_{X}\right\rangle= 10^{33}$\,erg\,s$^{-1}$ and $\langle L_{\text{opt}}\rangle=2\times 10^{31}$\,erg\,s$^{-1}$. On the contrary, for a cyclotron-dominant regime ($\epsilon_{s}\gg 1$), the optical luminosity can be larger than mean X-ray luminosity. We note that the equality between the two luminosities takes place for $\epsilon_{s} \sim 10$, and not for $\epsilon_{s} \sim 1$ as given by Eq.\,(\ref{epsilon_s}). The reason is that $\epsilon_{s}$ is evaluated just behind the shock and to achieve the equality between the mean luminosities in the whole column, cyclotron cooling must be more important than the case to achieve the equality only at the shock front.

These different cooling regimes depend on the polar parameters. Firstly, the mean X-ray luminosity $\langle L_{X} \rangle$ increases with the white dwarf mass and the accretion rate, which explains that the larger accretion rates present in intermediate polars induce larger X-ray luminosities than the ones observed in polars \citep{Warner1995}. Secondly, the mean optical luminosity, $\langle L_{\text{opt}} \rangle$, increases with the magnetic field. However, $\langle L_{\text{opt}} \rangle$ shows a maximum when the accretion rate increases. This maximum is due to the balance between the two radiative processes. Indeed, for a strong accretion rate, bremsstrahlung cooling is dominant. When the accretion rate decreases, the cyclotron cooling becomes more and more important until it dominates. In the same time, the available energy in the system decreases with the accretion rate. So although the cyclotron process dominates which implies that $\langle L_{X}\rangle \ll \langle L_{\text{opt}} \rangle$, the mean optical luminosities reaches a maximum at about $\epsilon_{s} \sim 3$ before decreasing for weaker accretion rates.

\section{QPOs as a function of the polar parameters}\label{section_3}

The balance between the two radiative processes determines the characteristics of the oscillations displayed by the luminosities. We present here a large parametric study aimed at exploring the influence of the polar parameters ($M_{\text{WD}}$, $B$ and $\dot{m}$) on the oscillation amplitudes and frequencies.

\subsection{Inconsistency between observations and simulations}

\begin{figure}
\centering
\includegraphics[width=9.3cm]{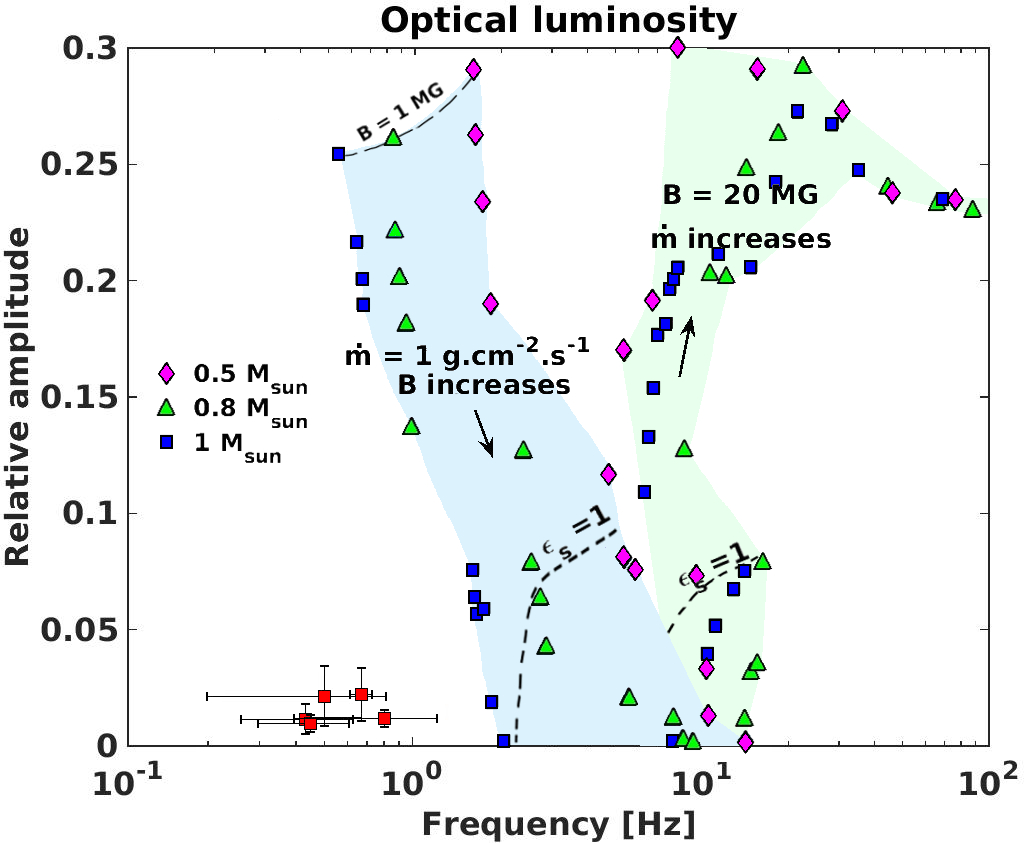}
\caption{Relative oscillation amplitudes of the optical luminosity as a function of the main frequencies for the models detailed in Table\,\ref{liste_simu}. The three white dwarf masses are displayed by diamond markers for $0.5\,M_{\odot}$, by triangle markers for $0.8\,M_{\odot}$ and by square markers for $1\,M_{\odot}$. The blue shaded region displays the range covered by the simulations with constant accretion rate ($\dot{m}=1$\,g\,cm$^{-2}$\,s$^{-1}$) and with a varying magnetic field (models (a), (b) and (c)). The magnetic field value starts at $1$ MG. The green shaded region shows the range for simulations with a constant magnetic field constant ($B=20$\,MG) and with a varying accretion rate (models (d),(e) and (f)). The  oscillations extracted from the RAMSES simulations are compared with observed QPOs displayed by the red points (Table\,\ref{QPO_data}).}\label{freq_ampO}
\end{figure}

\begin{figure}
\centering
\includegraphics[width=9.2cm]{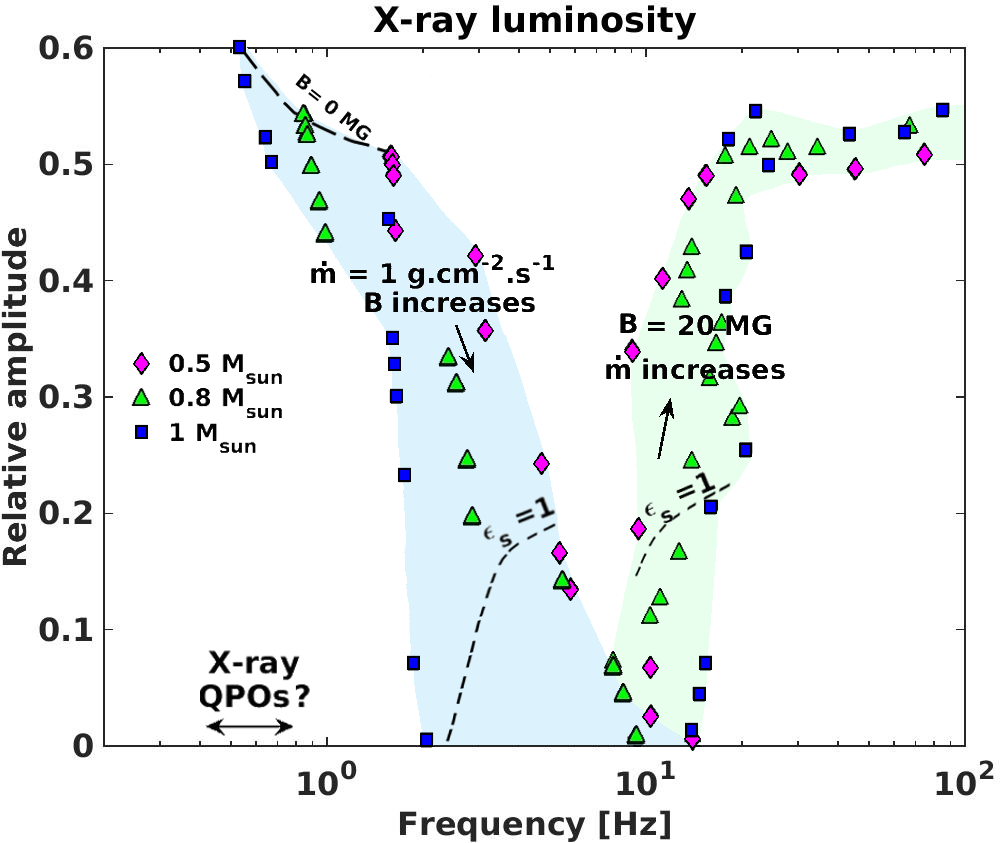}
\caption{Relative oscillation amplitudes of the X-ray luminosity as a function of the main frequencies for the models detailed in Table\,\ref{liste_simu}. The three white dwarf masses are displayed by diamond markers for $0.5\,M_{\odot}$, by triangle markers for $0.8\,M_{\odot}$ and by square markers for $1\,M_{\odot}$. The blue shaded region displays the range covered by the simulations with constant accretion rate ($\dot{m}=1$\,g\,cm$^{-2}$\,s$^{-1}$) and with a varying magnetic field (models (a), (b) and (c)). The magnetic field value starts at $0$ MG. The green shaded region shows the range for simulations with a constant magnetic field constant ($B=20$\,MG) and with a varying accretion rate (models (d),(e) and (f)). The oscillations extracted from the RAMSES simulations are compared with potential X-ray QPOs (see explanations in the text).}\label{freq_ampX}
\end{figure}

The relative amplitudes of oscillations derived from the optical and the X-ray luminosities are displayed, as a function of their frequencies, in respectively Fig.\,\ref{freq_ampO} and Fig.\,\ref{freq_ampX}. Each coloured point displays one simulation as detailed in Table\,\ref{liste_simu}. The blue shaded region displays the range covered by the simulations with constant accretion rate ($\dot{m}=1$\,g\,cm$^{-2}$\,s$^{-1}$), with a varying magnetic field ($B=0-22$\,MG) and with the three white dwarf masses ($M_{\text{WD}}=0.5\,M_{\odot}$ for model (a), $M_{\text{WD}}=0.8\,M_{\odot}$ for model (b) and  $M_{\text{WD}}=1\,M_{\odot}$ for model (c) in Table\,\ref{liste_simu}), while the green shaded region shows the range for simulations with a constant magnetic field ($B=20$\,MG), varying accretion rates ($\dot{m}=0.8-200$\,g\,cm$^{-2}$\,s$^{-1}$) and with the three white dwarf masses ($M_{\text{WD}}=0.5\,M_{\odot}$ for model (d), $M_{\text{WD}}=0.8\,M_{\odot}$ for model (e) and  $M_{\text{WD}}=1\,M_{\odot}$ for model (f) in Table\,\ref{liste_simu}). In the two regions, the three white dwarf masses are displayed by pink diamond markers for $0.5\,M_{\odot}$, by green triangle markers for $0.8\,M_{\odot}$ and by blue square markers for $1\,M_{\odot}$. In general, the results show that simulated oscillations extend over a frequency range of $f \sim 0.5-151$ Hz, with variation in the relative amplitude of the luminosity of $0.1-60$\% for X-rays and $0.1-30$\% for the optical emission.

Firstly, for the case of constant specific accretion rate ($\dot{m}=1$\,g\,cm$^{-2}$\,s$^{-1}$) and for the three white dwarf masses (blue region and models (a), (b) and (c)), the increase of the magnetic field leads to larger oscillation frequencies of both X-ray and optical luminosities. This is due to the link between the cooling characteristic time and the magnetic field as $f \sim 1/t_{\text{cool}} \sim B^{2.85}$ as displayed by \citet{Busschaert2015}. At the same time, we observe a drastic decrease of their relative amplitudes, as expected with the increase of cyclotron cooling. For all cases, the two luminosities are strongly coupled in frequency, as was illustrated in Fig.\,\ref{oscillations}, and the decrease of the relative amplitude is similar for both luminosities.
As the magnetic field is further increased, it reaches a critical value for which oscillations are completely inhibited by cyclotron cooling. The numerical limit of relative amplitudes is assumed at $10^{-3}$. Below this value, there is no longer oscillations. For white dwarf masses of $0.5$, $0.8$ and $1$\,M$_{\sun}$, these critical values are respectively $B=22$, $11$ and $5$\,MG. The damping of the oscillations for those masses corresponds to $\epsilon_{s} = 4.9$, $4.8$ and $0.8$. As mentioned earlier, these values of $\epsilon_{s}$ are different from $\sim 1$ because $\epsilon_{s}$ is evaluated just behind the shock and does not take into account the gradients induced by secondary shocks and complex wave structure in the accretion column.

Secondly, for a constant magnetic field ($B=20$ MG) and for the three white dwarf masses (green region and models (d),(e) and (f)), the increase of the specific accretion rate produces larger oscillation frequencies and relative amplitudes of both X-ray and optical luminosities. As the accretion rate increases, the dominant cooling process is first cyclotron emission and then bremsstrahlung. Consequently, for a given white dwarf mass and magnetic field, we find a critical accretion rate below which oscillations are inhibited. For white dwarf masses of $0.5$, $0.8$ and $1$\,M$_{\sun}$, the critical values of the specific accretion rate are respectively $0.8$, $3$ and $7$\,g\,cm$^{-2}$\,s$^{-1}$. The parameter $\epsilon_{s}$ is again different from one, its value being respectively $4.3$, $3.8$ and $3.1$. We notice that for relatively large specific accretion rates, above about $30$\,g\,cm$^{-2}$\,s$^{-1}$, the relative amplitudes of the oscillations seem to reach asymptotic values of $\sim 25$\% for the optical luminosities and $\sim 50$\% for the X-ray luminosities. The explanation of this asymptotic behaviour is still under study.

In both luminosities and both regions, the evolution of the oscillation frequencies shows different breaks which coincide with the apparition of a second and a third modes of oscillations. Indeed for weak magnetic fields and large accretion rates, the oscillation frequency evolves as the inverse of the cooling time, as described before, $f\sim1/t_{\text{cool}}$. For $\epsilon_{s}\sim 0.1-0.2$, there is a first break which corresponds to a second oscillation mode. Then a second break can appear which coincides with a third mode. Theses modes can be compared to those obtained by \citet{Busschaert2015}.
\newline

The characteristics of the simulated oscillations are compared with observed optical QPO values to investigate their origins. As explained previously, Table\,\ref{QPO_data} summarizes the parameters of the five polars showing QPOs in their light curves. The observed relative amplitude ($5$\,\%) of the optical luminosity and the QPO frequency ($f_{\text{QPO}} < 1$\,Hz) for those polars are shown in Fig.\,\ref{freq_ampO} by the red squares with their error bars. It is clear that the oscillations triggered by the radiative instability are incompatible with the observations. The parameter range of the simulations was chosen to cover the observed polar parameters showing QPOs. Indeed, we are able to recover the observed frequencies and relative amplitudes only by taking parameters outside the range of observed polars. For instance, we can recover an oscillation frequency of $f\sim 0.5-0.8$\,Hz with a relative amplitude of $\sim 10$\,\% only for a white dwarf mass of $M_{\text{WD}}=1.2$\,M$_{\odot}$, a very weak magnetic field $B\sim 2$\,MG and a specific accretion rate of $\dot{m}\sim 1$\,g\,cm$^{-2}$\,s$^{-1}$. Similar results can be obtained for a more common white dwarf mass $M_{\text{WD}}=0.4$\,M$_{\odot}$, a relatively weak magnetic field $\sim 4-5$\,MG and a low specific accretion rate $\dot{m}\sim0.1$\,g\,cm$^{-2}$\,s$^{-1}$. Again, these parameters are out of range of the observed polars and they would correspond to non magnetic or intermediate polars for which there are no QPO observations. 

Another important inconsistency is the presence in the simulations of oscillations in the X-ray luminosity, which are not detected by observations. Regarding the frequencies, we assume, based on previous results \citep{Busschaert2015} and the results presented in Fig.\,\ref{oscillations}, that the X-ray QPOs should oscillate at the same frequency as the optical QPOs. Analysis of observations reported in \citet{Bonnet2015} derived the upper limits of X-ray QPO amplitudes from the negative results of the \textit{XMM-Newton} observations for the five polars that we report in Table\,\ref{QPO_data}. We find that even for the object BL Hyi, which has the largest upper limit ($71$\%) and which would fulfil the QPO criterion in amplitude and frequency, the magnetic field value necessary in the simulations ($\sim 1$ MG) is not consistent with observations. Similarly, we find that for the other X-ray relative amplitude upper limits, the simulated oscillations would not match the observed frequencies.

\subsection{Discussion}

Although observations cannot directly determine the column cross-section, $S$, which is linked to the specific accretion rate by $\dot{m}=\dot{M}/S$, they provide some constraint on the fractional area that is covered by accretion. This is given by the ratio $S/S_{\text{WD}}$, where $S_{\text{WD}}$ is the white dwarf's surface. Values for this parameter range from $10^{-5}$ to $10^{-3}$ \citep{Wu1994, Cropper1999}, which for different white dwarf radii gives accretion cross sections in the range $S\sim 10^{13}-10^{16}$\,cm$^{2}$.
The cross-section appears in the cyclotron cooling function (see Eq.\,\ref{cycl}), and as a result it modifies the relative importance of cyclotron to bremsstrahlung radiative cooling. Thus the ratio of the characteristic cooling times, $\epsilon_{s}$, depends on the cross-section as:
\begin{equation}
\epsilon_{s}\sim ~ S^{1.425} ~ \dot{M}^{-1.85} ~  B^{2.85} ~  M_{\text{WD}}^{2.925} ~  R_{\text{WD}}^{-2.925}
\end{equation}
For a given white dwarf mass, magnetic field and accretion rate, an increase of the cross-section, and thus of the emitting volume, increases the cyclotron radiative losses. The effect is to decrease the amplitudes and frequencies of the oscillations in both X-ray and optical luminosities \citep{Mouchet2017}. On the contrary, decreasing the cross-section, and thus the parameter $\epsilon_s$ increases the bremsstrahlung radiative losses, which has the effect of producing higher frequency oscillations of larger amplitude. This makes the comparison with observations even more difficult. Indeed, previous work by \citet{Mouchet2017} has shown that changing the accretion cross section over three orders of magnitude was unable to reconcile the observed QPOs in V834 Centauri with the simulated oscillations. We remind that in this work, we have taken a constant and nominal value for the cross-section $S=10^{15}$\,cm$^{2}$ as in \citet{Busschaert2015}.

The inconsistency between the numerical predictions of optical QPOs with observations highlights the difficulty to understand their origins. Several possible processes in the accretion column could be at work to produce these oscillations and can explain our inconsistency but may not be revealed by oversimplified one-dimensional models.

Beyond the assumption of homogeneous accretion, several factors may induce the fragmentation of the accretion flow coming from the companion star towards the compact object \citep{Meintjes2004}. Besides, modelling of observations of the accretion columns (\citet{Potter2016} and references therein), find that it is necessary to consider more complicated geometries than a cylindrical homogeneous column in order to explain modulation of the observed flux \citep{Beuermann1987}, as for instance an extended arc-shaped region which may contain denser blobs \citep{Cropper1994}. The effects of clumpy accretion have been studied, for example, by a number of authors to explain the soft X-ray excess in polars \citep{Frank1988, Hameury1988, King2000, Schwarz2005}. 

Indeed the accretion column could also be fragmented by magnetohydrodynamical (MHD) effects, analogue to those simulated in Young Stellar Objects (YSO) \citep{Matsakos2013}. These two-dimensional simulations showed that the accretion column can break up into independently oscillating tubes (fibrils) which are radiatively unstable. For strong magnetic field, the global oscillation frequency can be suppressed if the fibrils are out of phase. Such a MHD mechanism could occur in polars. The coupling between the fibrils may induce an overall lower mean frequency in the luminosity than the one obtained in 1D simulations with homogeneous accretion flows as well as the decrease in the overall QPO amplitudes. However, polars display much stronger magnetic fields and accretion rates than the ones in YSO ($B_{\textrm{YSO}}\sim1$\,kG and $\dot{m}_{\textrm{YSO}}\sim10^{-4}$\,g\,cm$^{-2}$\,s$^{-1}$). Indeed, the white dwarf magnetic fields impose a much smaller plasma-$\beta$ than YSOs in the post-shock region: $\beta_{\text{polar}}={P}/({B^{2}/8\pi})\sim 10^{-4}-10^{-3}\ll \beta_{\textrm{YSO}} \sim10^{-1}$. Besides, the non-linear regime of the accretion column and in particular the secondary shock dynamics may substantially modify the dynamics of the fibrils in polars. New interesting investigations on multi-dimensional effects are promising and may help to understand the observational data. They also indicate the need to move beyond one-dimensional simulations. 
  
	The radiative and dynamic feedback of the white dwarf photosphere on the accretion column may also be important. For example, reprocessed radiation from the photosphere may heat the bottom of the accretion column, leading to an increase of the local sound speed and preventing the formation of the acoustic horizons. This should lead to the suppression of the secondary shock, which as shown in our work, largely determines the shock oscillation dynamics. Consequently, an accurate modelling of the photosphere may also be necessary in the future to disentangle the physics of accretion dynamics in polars. Furthermore, the behaviour and stability at the interface between the degenerate matter coming from the white dwarf and non-degenerate matter from the accretion column has not been explored yet and may lead to new interesting phenomena.
    
Finally, the large incertitude on the masses of white dwarfs and accretion rates, calls into question the use of fluid models in certain regimes. For example, for large white dwarf masses and magnetic fields, or for small accretion rates, two-temperature or collisionless effects on the oscillations have not been either explored.

\section{Conclusion}

We have shown that there is a major inconsistency to interpret QPOs in terms of the accretion shock oscillations model. Indeed, matching the observed QPOs with simulations is not possible within the current model of accretion columns.

In this work, we present simplified semi-analytical solutions of the steady-state equations of the standard model of accretion columns in polars and new one-dimensional hydrodynamic simulations of accretion shocks performed with the RAMSES code with gravity and radiative cooling. The steady-state approximation enables us to determine the radiative regime of accretion,  which influences for example the height of the shock or the impact of the white dwarf gravity on the column stratification. In particular, it allows us to determine the position of the acoustic horizon where secondary shocks are expected to form.

To move beyond the steady-state approximation we performed RAMSES simulations including gravity, cyclotron and bremsstrahlung radiative losses, for a wide range of parameters. We have demonstrated that shock oscillations are the result of the interplay of complex shock dynamics triggered by the cooling and the Falle radiative instabilities. Specifically, the secondary shock is formed at the acoustic horizon in the post-shock region in good agreement with our estimates from the steady-state solutions. We demonstrate that the secondary shock is essential to sustain the oscillations of the accretion shock, at the average height predicted by steady-state accretion models. Without it, we expect the accretion shock to collapse on the white dwarf photosphere, as observed in simulations of YSO where no secondary shocks are formed.

To compare simulations with observations, synthetic optical and X-ray luminosities are computed, as well as an extensive Fourier analysis of the oscillation amplitudes and frequencies extracted from simulations. The simulated parameter space covers all the polars showing optical QPOs. The results are compared with observed optical QPOs as well as with upper limits of X-ray QPOs. We find a strong inconsistency between the simulated and observed QPOs amplitudes and frequencies. To match the observed values, would require a combination of polar parameters, such as mass, accretion rate and magnetic field, inconsistent with the observed parameters. At best, the matching parameters would correspond to non-magnetic and intermediate polars. 

Within the framework presented here, we find no possibility to reconcile simulations with observations. This difficulty highlights the limits of one-dimensional simulations to study the dynamics of accretion columns, suggesting that multi-dimensional simulations are needed to understand the non-linear dynamics of accretion columns in polars and the origins of QPOs.

\section*{Acknowledgements}
Part of this work was supported by the "Programme National de
Physique Stellaire" (PNPS) of CNRS/INSU, France. This work was also partly supported by the ANR Blanc grant no. 12-BS09-025-01 SILAMPA and the LABEX Plas@par grant no. 11-IDEX-0004-02 from the French agency ANR.



\bibliographystyle{mnras}
\bibliography{biblio}




\bsp	
\label{lastpage}
\end{document}